\documentclass[aps,pra,twocolumn,showpacs,floatfix]{revtex4}

\usepackage{amsmath}
\usepackage{amssymb}
\usepackage{graphicx}
\usepackage{epstopdf}
\usepackage{verbatim}

\begin{document}

\newcommand{\bmath}{\begin{displaymath}}
\newcommand{\emath}{\end{displaymath}}

\newcommand{\be}{\begin{equation}}
\newcommand{\ee}{\end{equation}}
\newcommand{\bea}{\begin{eqnarray}}
\newcommand{\eea}{\end{eqnarray}}
\newcommand{\non}{\nonumber\\}
\newcommand{\bmultl}{\begin{multline}}
\newcommand{\emultl}{\end{multline}}

\newcommand{\bsubeq}{\begin{subequations}}
\newcommand{\esubeq}{\end{subequations}}
\newcommand{\bitemize}{\begin{itemize}}
\newcommand{\eitemize}{\end{itemize}}
\newcommand{\ket}[1]{\left|{#1}\right\rangle}
\newcommand{\bra}[1]{\left\langle{#1}\right|}
\newcommand{\abs}[1]{\left|{#1}\right|}
\newcommand{\re}{\mathrm{Re}}
\newcommand{\im}{\mathrm{Im}}
\newcommand{\bmx}{\begin{bmatrix}}
\newcommand{\emx}{\end{bmatrix}}
\newcommand{\bsmx}{\begin{smallmatrix}}
\newcommand{\esmx}{\end{smallmatrix}}

\def\ud{\mathrm{d}}
\def\dt{\frac{\partial}{\partial t}}
\def\R{\vec{\hat{R}}}
\renewcommand{\Re}{\mathrm{Re}}
\renewcommand{\vec}[1]{\underline{#1}}
\newcommand{\mat}[1]{\mathbf{#1}}
\newcommand{\rev}[1]{\vec{r}^{(#1)}}
\newcommand{\lev}[1]{\vec{l}^{(#1)}}

\def\wR{\omega_R}
\def\da{\delta{}a}
\def\db{\delta{}b}
\def\dc{\delta{}c}

\newcommand{\bquote}{\quotedblbase{}}
\newcommand{\equote}{\textquotedblright{ }}

\title{On the critical exponent of a quantum noise driven phase transition: the open system Dicke-model}
\author{D. Nagy}
\author{G. Szirmai}
\author{P. Domokos}

\affiliation{Research Institute for Solid State Physics and Optics, H-1525 Budapest P.O. Box 49, Hungary}
\begin{abstract}
The quantum phase transition of the Dicke-model has been observed recently in a system formed by motional excitations of a laser-driven Bose--Einstein condensate coupled to an optical cavity \cite{Baumann2010Dicke}. The cavity-based system is intrinsically open: photons can leak out of the cavity where they are detected. Even at zero temperature, the continuous weak measurement of the photon number leads to an irreversible dynamics towards a steady-state which exhibits a dynamical quantum phase transition. However, whereas the critical point and the mean field is only slightly modified with respect to the phase transition in the ground state, the entanglement and the critical exponents of the singular quantum correlations are significantly different  in the two cases.
\end{abstract}


\maketitle

\section{Introduction}

Experiments with ultracold atomic gases in optical fields laid down a new path to discover strongly correlated many-body quantum systems. In particular, the high degree of control over the interaction parameters allows for using atomic systems as quantum simulators of generic theoretical models \cite{Jordens2010Quantitative}. Central to these efforts lies the possibility of observing quantum phase transitions (QPT).  At effectively zero temperature ($T=0$), by tuning an external field acting on the system, it can be scanned through a quantum critical point which separates regions with different symmetries in the ground state. One celebrated example is the QPT from a superfluid to a Mott insulator in the Bose-Hubbard model \cite{Jaksch1998Cold} that was realized with a gas of ultracold atoms in an optical lattice \cite{Greiner2002Quantum}. Additional quantum phases appear in this system when dipole-dipole interaction is present \cite{Goral2002Quantum}. 

A fundamental question is how quantum phase transitions are influenced by non-equilibrium conditions. The  ordinary way to prepare a stationary system out of equilibrium at $T=0$ can be illustrated by a BEC in a rotating trap. It undergoes  the vortex formation QPT  above a critical angular velocity \cite{Dagnino2009Vortex}. External driving can impose that only a certain subset of states in the Hilbert space, those having a given moment of inertia in the previous example, be populated. Similar effect has been described for a spin chain in ring geometry:  it can manifest criticality while being confined into the subspace of energy current carrying states \cite{Antal1997Nonequilibrium}. In both examples the system is effectively Hamiltonian.  

One can go beyond the effectively Hamiltonian systems by adding external non-equilibrium noise on critical states. It was shown that  the $1/f$ noise, ubiqitous in electronic circuits, preserves the quantum phase transition in the steady state of a system, moreover, it gives a knob to tune  the critical exponent by the noise strength \cite{DallaTorre2010Quantum}.  This is in sharp contrast with the well-known effect of thermal fluctuations that destroy quantum critical correlations. In a more general level, {\it reservoir engineering} is a route towards designing specific noise sources in a dissipation process which leads to pure many-body states in the dynamical steady state. An example is a lattice gas immersed in a BEC of another species of atoms \cite{Diehl2008Quantum}, which serves as a zero-temperature reservoir of Bogoliubov excitations. The resulting dissipative Bose-Hubbard model exhibits a dynamical phase transition between a pure superfluid state and a thermal-like mixed state as the on-site interaction is increased \cite{Diehl2010Dynamical}. Note that this method for the preparation of strongly correlated quantum states makes dissipation to be a resource for quantum simulation \cite{Weimer2010Rydberg} and universal quantum computation \cite{Verstraete2009Quantum}.

In this paper we will consider the bare electromagnetic vacuum at $T=0$ as a reservoir and its effect on a Dicke-type Hamiltonian system which is known to produce a singularity of the ground state \cite{Hillery1985Semiclassical}. Placed into a dissipative environment, the system evolves irreversibly into a steady state which is a dynamical equilibrium between driving and damping. The intrinsic noise accompanying the dissipation process is in accordance with the dissipation-fluctuation theorem. Even in this very natural case of non-equilibrium, the loss does not destroy quantum criticality.  But what is the relation of the criticality expected in the steady-state to that of the ground state in the closed Hamiltonian system? 

\begin{figure}[ht!]
\centering
\includegraphics[width=\columnwidth]{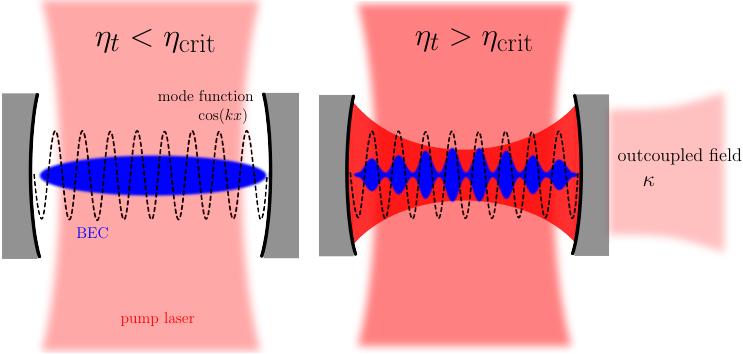}
\caption{Self-organization phase transition of a BEC in a cavity. Below a threshold in the transverse driving field (left) the condensate is quasi-homogeneous, and there are no photons inside the cavity. Above threshold (right), a standing matter wave of period $\lambda$ appears that scatter photons into the cavity.}
\label{fig:scheme}
\end{figure}

Our specific example is the self-organization phase transition of laser-driven atoms in an optical resonator \cite{Domokos2002Collective,Asboth2005Selforganization,Black2003Observation,Keeling2010Collective,Gopalakrishnan2009Emergent,Vidal2010Quantum}. The laser impinges on the atoms from a direction perpendicular to the resonator axis (see Fig.~\ref{fig:scheme}). Below a critical value of the pump intensity, the spatial distribution of the atoms is homogeneous along the axis and the mean cavity photon number is zero, since the photons scattered by the atoms into the cavity interfere destructively. Above a threshold pump power, there appears a wavelength-periodic modulation of the distribution, from which laser photons can be Bragg-scattered into the resonator. Spontaneous symmetry breaking takes place between two possible solutions for the cavity field phase and the atomic distribution. This is a non-equilibrium phase transition, which has an experimentally accessible $T=0$ limit if the atomic cloud is represented by a Bose-Einstein condensate. The phase diagram has been experimentally mapped by Baumann et al.\ \cite{Baumann2010Dicke}.

\section{Open system description}

Consider the dispersive coupling of a one-dimensional matter wave field $\Psi(x)$ to a single cavity mode $a$ in the transverse pump geometry shown in Fig.~\ref{fig:scheme}. The dispersive limit appears when the laser pump is far detuned from the atomic resonance ($\Delta_A = \omega - \omega_A$ exceeds the atomic linewidth $\gamma$ by orders of magnitude). In the frame rotating at the pump frequency $\omega$,  the many-particle Hamiltonian reads
\begin{multline}
\label{eq:H_total}
H/\hbar = -\Delta_C\,a^\dagger{}a 
 + \int_0^L \Psi^\dagger(x)\bigg[-\frac{\hbar}{2\,m}\frac{d^2}{dx^2}  \\
  + U_0\,a^\dagger{}a\cos^2(kx) + i \eta_t \cos{kx} (a^\dagger -a)\bigg]\Psi(x)dx\,.
\end{multline}
The detuning $\Delta_C = \omega - \omega_C$ defines the effective photon energy in the cavity. Atom-atom s-wave collisions are neglected,  the length of the condensate along the cavity axis is $L$.  The atom--light interaction originates from coherent photon scattering. The absorption of a cavity photon and stimulated emission back into the cavity gives rise to the term proportional to $U_0 = g^2/\Delta_A$. The coherent redistribution of photons between the pumping laser and the cavity mode results in an effective pump with amplitude $\eta_t = \Omega{}g/\Delta_A$. Note that this term describes the external driving of the system, and the explicit time-dependence in the optical frequency range, due to the laser field, has been eliminated by the transformation into the rotating frame.
The remaining frequencies are in the kHz range of the recoil frequency $\wR= \hbar k^2/2m$.

The critical behaviour can be described in a subspace spanned by two motional modes, i.e., 
\be
\label{eq:state_ansatz}
\Psi(x) = \frac{1}{\sqrt{L}} c_0 + \sqrt{\frac{2}{L}} c_1 \cos{k x}\,,
\ee
with the bosonic annihilation operators  $c_0$ and $c_1$. With the closed subspace constraint $c_0^\dagger c_0 + c_1^\dagger c_1 =N$ imposed, 
the Hamiltonian of the system formally reduces to that of the Dicke model \cite{Nagy2010DickeModel}. Originally, it was introduced to describe the dipole coupling of $N$ two level atoms to a single quantized field mode \cite{Dicke1954Coherence}.  It is known for a long time that the Dicke model can exhibit a thermodynamic phase transition at finite temperature \cite{Hepp1973Equilibrium}, and a quantum phase transition at zero temperature \cite{Hillery1985Semiclassical} between an unexcited normal phase and a superradiant phase, where both the atoms and the mode are macroscopically excited. There is a renewed interest in studying the zero-temperature properties of this system with particular respect to critical entanglement \cite{Lambert2004Entanglement}, finite-size scaling \cite{Vidal2006Finitesize,Liu2009LargeN} and quantum chaos \cite{Emary2003Quantum}.  Dimer et al.\ proposed a realization of the Dicke model with photon loss by means of multilevel atoms coupled to a ring cavity mode via Raman transitions \cite{Dimer2007Proposed}. Another collective spin model exhibiting dynamical QPT, the Lipkin-Meshkov-Glick model was constructed in cavity QED systems \cite{Morrison2008Dynamical}. 

We consider a single dissipation channel which is the photon leakage through one of the mirrors. The corresponding dissipation process can be modeled by a Heisenberg-Langevin equation for the field amplitude $a$, which includes a loss term with rate $\kappa$ and a Gaussian noise operator $\xi(t)$,
\be
\label{eq:loss}
\frac{d}{dt}a = -i [a\,,H] -\kappa{}a + \xi\,.
\ee
The effect of continuous weak measurement of the photon number is described by the same equation. The noise operator $\xi$ has zero mean and its only non-vanishing correlation is $\langle \xi(t) \xi^\dagger(t')\rangle = 2 \kappa \delta(t-t')$ at $T=0$. For finite temperature, other correlations would appear proportional with the thermal photon number. The given second-order correlation expresses the fluctuation-dissipation theorem. The noise operator can  be seen as a necessary source for maintaining the commutation relation and general algebraic properties during the time evolution.

This equation applies to the decay of a single uncoupled harmonic oscillator. However, in the present case, the  field mode interacts with the matter wave field. The decay process takes place in the optical frequency range which is many orders of magnitude above the characteristic frequencies of the  interaction (comparable to the recoil frequency in the kHz range). Therefore, the coupling to the other atoms has a only a negligible influence on the decay of the photon mode and Eq.~(\ref{eq:loss}) holds for the interacting system \cite{Zoubi2003Dissipations}. But one must keep in mind that the original problem is intrinsically time-dependent,  and thus there is an energy current from the laser into the reservoir through the system. 

\section{Steady state vs. ground state}

\begin{figure}[ht!]
\centering
\includegraphics[width=0.89\columnwidth]{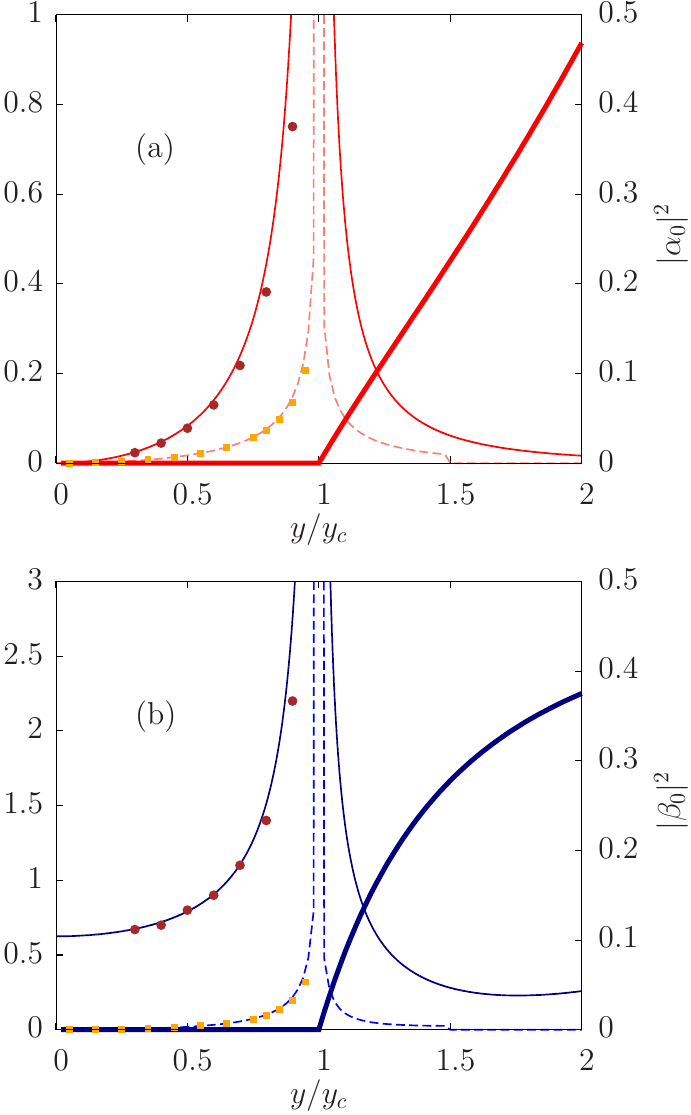}
\caption{Comparison of the criticality in the ground state of the closed, and in the steady-state of the open Dicke models.  The mean fields and the incoherent excitation numbers of the
  photon (a) and the atom (b) fields are plotted as a function of the coupling normalized to its critical value $y_c$. With this scaling, the mean field amplitudes (thick
  solid lines) coincide in the two states.  There is a marked difference between the excitation numbers
  in the steady state (thin solid lines) and in the ground state (thin  dashed lines). The points show the numerical results for finite atom
  number, which we obtained using a general quantum simulation framework for cavity QED systems \cite{Vukics2007CQED}. The steady state (circles) and the ground state (squares) are calculated for $N=200$ and $400$, respectively. The parameters are $\delta_C=-2$, $\kappa=2$  $\omega_R$.}
\label{fig:comp}
\end{figure}
Taking the $N\rightarrow\infty$ limit, the mean-field approach, which can be adopted both in the lossy and lossless cases of the Dicke-model (see Appendix A), gives an adequate approximation of the true steady-state and ground state, respectively.  Fig.~\ref{fig:comp} presents a comparison of the quantum critical behavior in these two cases. The operators are split
into their steady-state expectation values and quantum fluctuations,
\be
\label{eq:sep}
a = \sqrt{N}\alpha_0 + \delta{}a\;;\quad c_k =
e^{-i\mu{}t}(\sqrt{N}\gamma_k + \delta{}c_k)\,, 
\ee 
where $\alpha_0$ corresponds to a coherent state in the cavity, and $\gamma_0$, $\gamma_1$ are the condensate wave function components. The trivial part of the time evolution driven by the chemical potential $\mu$ is separated from the atomic operators. We can choose  $\gamma_0 = \sqrt{1 - \beta_0^2}$ and $\gamma_1 = \beta_0$, where $0 \leq \beta_0 \leq 1$.  The mean-field amplitudes $|\alpha_0|^2$,  $\beta_0^2$  are plotted by thick lines (right scale) against the pumping strength $y=\sqrt{2 N}\, \eta_t$ normalized to the critical value $y_c$.  The critical point separates two phases with different symmetries. In the normal phase $\alpha_0 = \beta_0 = 0$, in the superradiant phase there are two mean-field solutions corresponding to $\pm \alpha_0$ and $\gamma_1 = \pm\beta_0$. The critical point $y_c= $ depends on $\kappa$. However, the mean-field solutions $\alpha_0$ and $\beta$  for the steady-state and for the ground state, expressed as a function of $y/y_c$,  overlap.  The perfect overlap is not a general property, but there is always a smooth connection so that  they coincide for $\kappa\rightarrow 0$. This is because the mean-field solution is dominated by the identical Hamiltonian part of the dynamics. In contrast, the correlation functions of quantum fluctuations may signify different kind of  second-order phase transitions, because the quantum noise source in the open system is dissipation (or the back-action of the measurement), which is completely missing from the closed system.

\begin{figure}[ht]
\centering
\includegraphics[width=0.66\columnwidth]{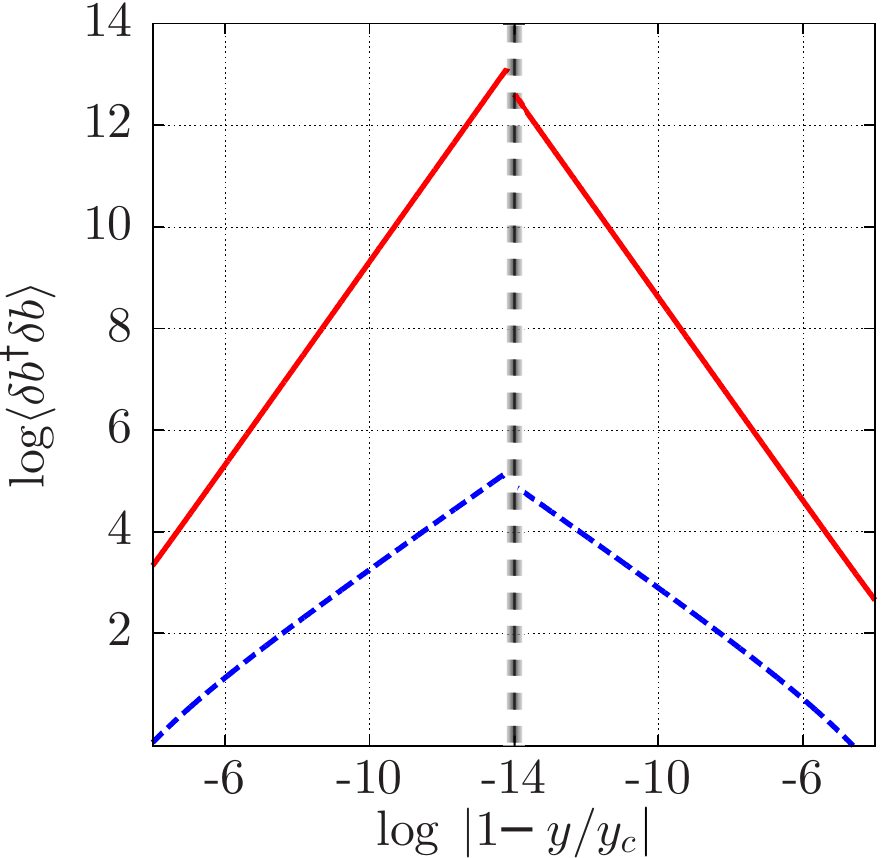}
\caption{Critical exponents for the phase transition in ground state and in the steady state. The scaling of the condensate depletion with the relative
  deviation from the critical point is shown in a log-log scale on both sides of the critical point. At the center $|1 - y/y_c| =
  e^{-14}$. We compare the averages in the steady state (solid red lines) for $\kappa = 2\wR$ and in
  the ground state of the system (dashed blue lines) without dissipation
  ($\kappa = 0$). The corresponding exponents are $-1.00$ and $-0.50$,
  respectively. Other parametes are $\delta_C=-2$.}
\label{fig:exponent}
\end{figure}

\begin{figure}
\centering
\includegraphics[width=0.66\columnwidth]{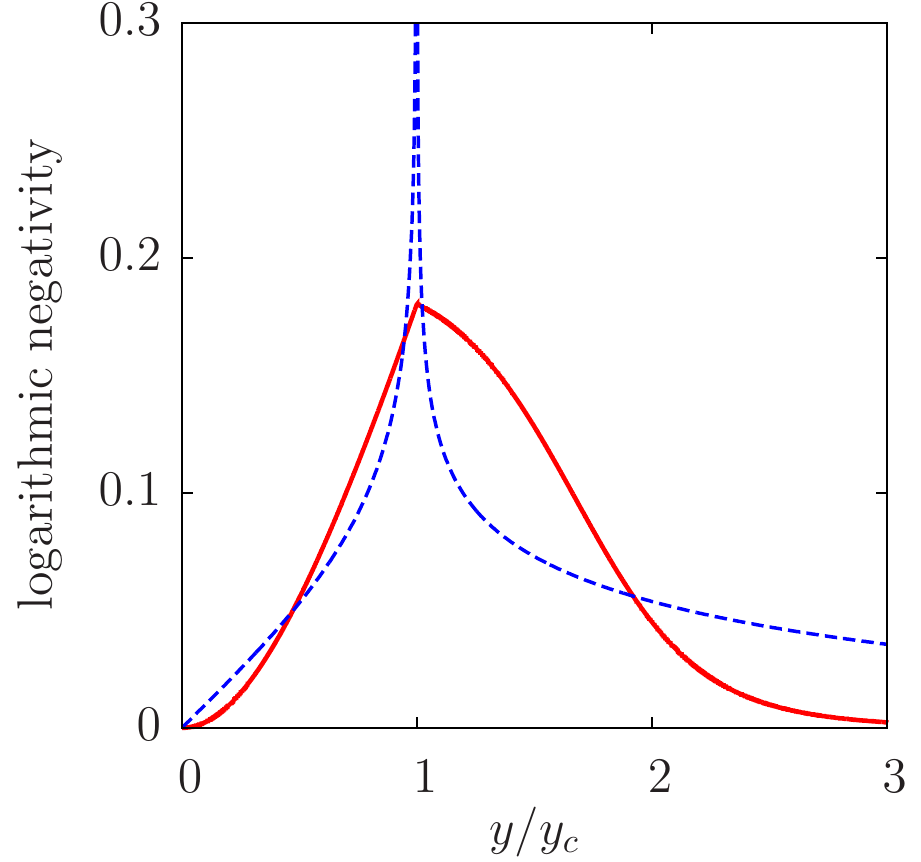}
\caption{Entanglement at the critical point. Logarithmic negativity as a function of the transverse pumping strength $y$ for the steady-state (solid red lines, $\kappa = 2$ $\wR$) and for the ground state (dashed blue lines, $\kappa = 0$). Parameters: $\delta_C = -2$.}
\label{fig:logneg_ss}
\end{figure}

For the matter field we consider  the excitation mode $\delta{}b$, which is orthogonal to the condensate wave function. In the normal phase, $\delta{}b = \delta{}c_1$, while in the superradiant phase $\delta{}b = -\beta_0{}\delta{}c_0 + \sqrt{1 - \beta_0^2}\delta{}c_1$. The figure shows the number of incoherent photons $\langle\da^\dagger\da\rangle$ (left scale, thin solid red line) and the \emph{condensate depletion} $\langle\db^\dagger\db\rangle$ (thin solid blue line). The same physical quantities are plotted with thin dashed lines. There is a divergence for both cases, however, the exponent is $-0.5$ for the ground state (usual mean-field exponent) whereas it is  $-1$  for the the steady state (see log-log scale in Fig.~\ref{fig:exponent}). 
This difference is independent of $\kappa$ and is related to the different physical origins of the
divergence. On the one hand, the
ground state is a two-mode squeezed state  with a squeezing parameter which tends to infinity at the
critical point. This results in a singularity of the entanglement between the
cavity and the atomic subsystems. On the other hand, the steady state is driven by 
quantum noise associated with dissipation (or measurement), which heats up the quasi-normal mode population infinitely where the imaginary part of its eigenvalue vanishes. The steady-state 
is a mixed state having a regular entanglement at the critical
point, reflected by the logarithmic negativity $E_{\cal N}$ in Fig.~\ref{fig:logneg_ss}. 

The steady-state divergence occurs, as mentioned above, when one of the quasi-normal modes has zero damping. This means a critical slowing down in reaching the steady-state. Therefore, in an experiment where the system is typically launched from the quasi-adiabatically tuned ground state, the photon number generated in the cavity during a finite time is expected in between that of the ground state and the steady state. Below threshold, where the mean field vanishes, any detected photon corresponds to the fluctuations $\langle\da^\dagger\da\rangle$, and a measurement above the ground state level indicates photons generated by the quantum noise penetrating into the cavity.

\section{Discussion}

In this paper we have adapted the famous Dicke-model to an intrinsically \emph{non-equilibrium} setting and pointed out distinctive features of this experimentally accessible driven-damped open system.  The underlying Dicke problem, a closed, conservative system,  has been a subject of intensive research for many decades. It remained an intriguing question, however, what happens with the critical point under non-equilibrium effects? While in classical physics the extension from equilibrium to non-equilibrium systems has been extensively studied, this step has not been made in quantum theory.
On the other hand, the ongoing experimental work will significantly shape the research on quantum phase transitions, too. In particular, we believe that the cavity QED based systems given in several laboratories worldwide raise relevant new aspects for classifying quantum critical phenomena. In this paper we revealed non-equilibrium critical effects in the case of the simplest possible environment.  It consists of a single well-defined dissipation channel (photon leakage out of the cavity), which is equivalent with the back-action of a weak quantum measurement on the photon number observable (continuous photo-detection).  This innocent looking intrusion in the system drastically modifies the critical exponent of the singularity at the phase transition point.

\section{Acknowledgements}

This work was supported by the Hungarian National Office for Research and Technology under the contract ERC\_HU\_09 OPTOMECH, the Hungarian National Research Fund (OTKA T077629) and the Hungarian Academy of Sciences  (Lend\"ulet Program, LP2011-016).

\appendix

\section{Mean-field solution of the two-mode model}

We solve the steady state of the system within a mean-field approach, which is consistent with the assumptions that (i) there is a macroscopically populated BEC wave function, and (ii) the state of the cavity field is close to a coherent state. After restricting the atomic dynamics to the spatial modes Eq. (2) (see also Ref. \cite{Nagy2010DickeModel}), we proceed by obtaining the equations of motion of the operators $a$, $c_0$, $c_1$. The first equation is the Heisenberg--Langevin equation given by Eq. (3), while the Heisenberg equations of motion for the operators $c_i$ follow directly from the Hamiltonian dynamics provided by Eq. (1).
\begin{subequations}
\label{eq:motion}
\be
\dot{a} = \left[ i\left(\delta_C - \frac{u}{N}c_1^\dagger{}c_1\right) - \kappa \right] a + \frac{y}{2\sqrt{N}}\left(c_0^\dagger c_1 + c_1^\dagger c_0 \right) + \xi \,,
\ee
\be
\dot{c}_0 = i\left[\frac{\wR}{2} + \frac{u}{2N} a^\dagger{}a\right] c_0 +  \frac{y}{2\sqrt{N}} \left(a^\dagger - a\right) c_1 \,,
\ee
\be
\dot{c}_1 = -i\left[\frac{\wR}{2} + \frac{u}{2N} a^\dagger{}a\right] c_1 +  \frac{y}{2\sqrt{N}} \left(a^\dagger - a\right) c_0 \,,
\ee
\end{subequations}
where we introduce the parameters $\delta_C = \Delta_C - 2u$, $\wR = \hbar{}k^2/(2m)$, $u = NU_0/4$ and $y = \sqrt{2N}\eta_t$. Note that in the thermodynamic limit $N\rightarrow \infty$ and $U_0, \eta_t \rightarrow 0$, while $u$ and $y$ are kept constant, and they can be expressed with the atom density. The noise operator $\xi$ has zero mean and its only non-vanishing correlation is $\langle \xi(t) \xi^\dagger(t')\rangle = 2 \kappa \delta(t-t')$, with $2\kappa$ being the photon loss rate \cite{Gardiner1985Input,Louisell1990Quantum}.

By using the decomposition of the operators to mean part and fluctuations, given by Eq. (4), in the equations of motion \eqref{eq:motion} and neglecting fluctuations we arrive to the mean-field equations determining $\alpha_0$, $\gamma_0$ and $\gamma_1$. Since the BEC wave function is normalized to unity, we can choose $\gamma_0 = \sqrt{1 - \beta_0^2}$ and $\gamma_1 = \beta_0$, where $0 \leq \beta_0 \leq 1$. By choosing $\beta_0$
positive, we select one of the two mean field solutions, that also
fixes the phase of the cavity field. For the steady state, we obtain
\begin{subequations}
\label{eq:cond}
\be \left[i\left(\delta_C - u\beta_0^2\right) - \kappa\right] \alpha_0
= -y \beta_0 \sqrt{1-\beta_0^2}\,,
\ee 
\be
\left(\wR + u |\alpha_0|^2 \right)\beta_0 = -y\;\mathrm{Im}(\alpha_0) \frac{1 - 2
  \beta_0^2}{\sqrt{1  - \beta_0^2}}\,,
\ee 
and the chemical potential $\mu = - \frac12
(\wR + u |\alpha_0|^2)/(1 - 2\beta_0^2)$.
\end{subequations}

Note that the coherent field amplitude $\alpha_0$ is complex, while
$\beta_0$ is real. The solution $\alpha_0 = \beta_0 = 0$ always
satisfies these equations, and it corresponds to the normal phase in
which the condensate is homogeneous ($\gamma_0 = 1$) and there is no
photon inside the cavity. Above the pumping threshold, $y_{c}^2 = - \wR
({\delta_C^2 + \kappa^2})/{\delta_C}$, the solution bifurcates, and the normal phase looses stability. The stable solution becomes 
\be
\label{eq:beta0}
\beta_0^2 = \frac{\delta_C}{u}\left(1 - \sqrt{1 -
  \frac{u}{\delta_C}\frac{y^2 - y_c^2}{y^2 + u \wR}} \right)\,,
\ee
which corresponds to the superradiant phase, where the condensate is
modulated ($\beta_0 > 0$) and the cavity field is finite
($|\alpha_0|^2 > 0$).  For $u=0$, the expression in
Eq.~(\ref{eq:beta0}) needs to be reformulated as $\beta_0^2 = (y^2 -
y_c^2)/2y^2$, and accordingly the mean field amplitudes coincide
both for the steady state and for the ground state ($\kappa=0$), if 
expressed as a function of $y/y_c$, leaving $\kappa$ the only role of
shifting the critical pumping strength $y_c$. 

The critical behaviour is unaffected by $u$, and the parameters $u$
and $y$ can be tuned independently, thus for simplicity we set
$u=0$ in the discussion of this Letter.

\section{Fluctuations in the steady state}

To go beyond mean field one has to keep the operator valued fluctuations $\delta a$ and $\delta c_i$ in Eqs. \eqref{eq:motion}. We consider quantum fluctuations up to linear order. Note, that the zeroth order term vanishes due to the mean field equations, and we arrive to a set of linear, stochastic differential equations for the fluctuations.
There are two types of fluctuations in the atom field $(\delta{}c_0$, $\delta{}c_1)$. The zero mode fluctuations, $\delta{}c = \sqrt{1 - \beta_0^2}\delta{}c_0 + \beta_0\delta{}c_1$, give rise to a phase diffusion of the condensate. The dynamics of the zero mode decouples from that of the other types of fluctuations. We are interested in the dynamics of the non-zero mode $\delta{}b = -\beta_0{}\delta{}c_0 + \sqrt{1 - \beta_0^2 }\delta{}c_1$, which describe the condensate depletion $\delta{}N = \langle \delta{}b^\dagger\delta{}b\rangle$. The coupled equations of motion read
\begin{subequations}
\label{eq:fluct}
\begin{multline}
\frac{d}{dt}\da = \left[ i\left(\delta_C - u\beta_0^2\right) -
  \kappa \right] \da + \xi \\+ \left[\frac{y}{2}(1 - 2\beta_0^2) - 
  i u \alpha_0\beta_0 \sqrt{1 - \beta_0^2}\right] (\db^\dagger +\db)\,,
\end{multline}
\begin{multline}
\frac{d}{dt}\db = -i\,\frac{\wR + u |\alpha_0|^2}{1 - 2\beta_0^2}\,\db 
  + \frac{y}{2}(1 - 2\beta_0^2)\,(\da^\dagger - \da) \\
  -i u \beta_0 \sqrt{1 - \beta_0^2}\,(\alpha_0\da^\dagger + \alpha_0^{*}\da)\,, 
\end{multline}
\end{subequations}

\begin{figure}
\centering
\includegraphics[width=0.9\columnwidth]{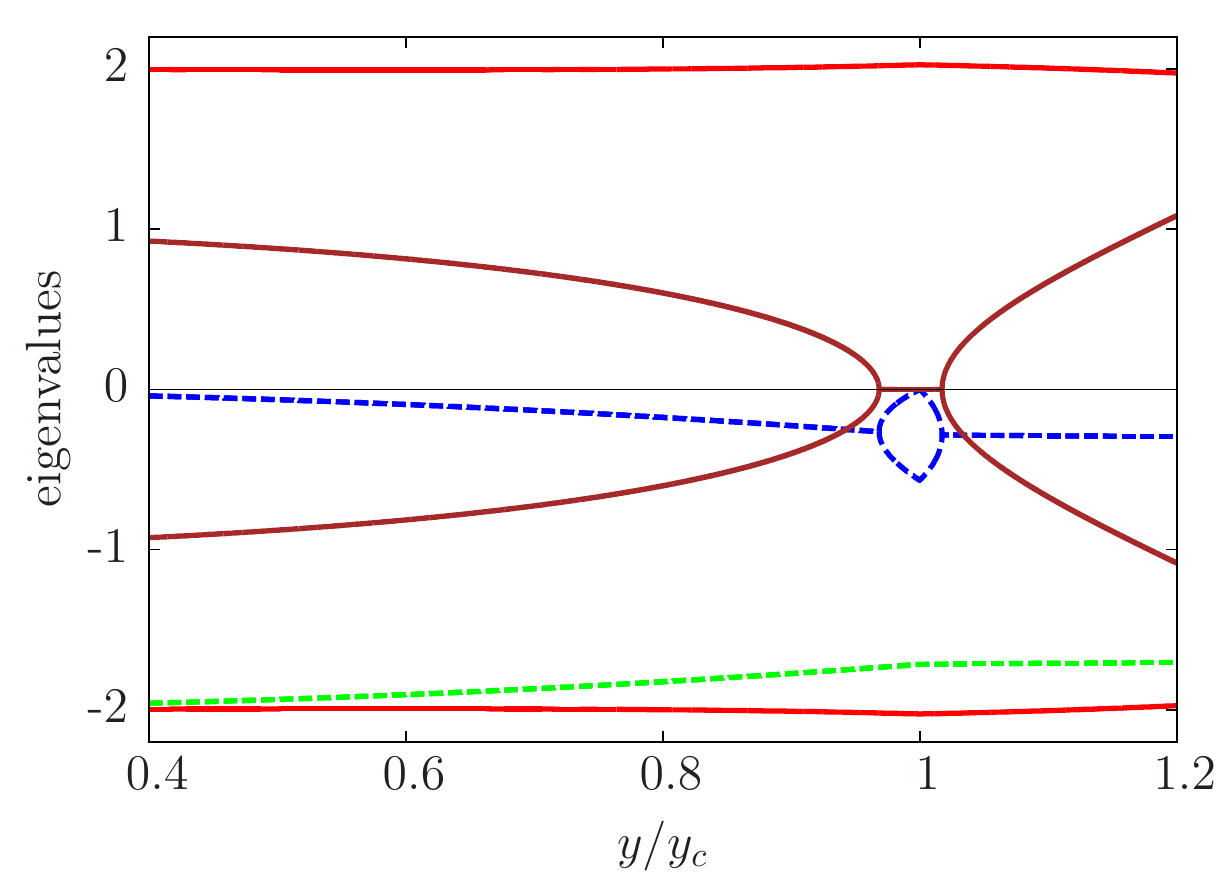}
\caption{Spectrum of the linear stability matrix $\mat{M}$ vs the
  transverse pump strength $y$. Solid lines (dashed lines) correaspond
  to the imaginary (real) part of the eigenvalues. The real parts are
  the same for a complex conjugate pair. Parameters: $\delta_C = -2$,
  $\kappa = 2$ $\wR$.}
\label{fig:evalues}
\end{figure}

We solve Eqs.~(\ref{eq:fluct}a-b) by
calculating the normal mode excitations of the system. 
Arranging the fluctuations in the vector
$\R=[\da,\da^\dagger,\db,\db^\dagger]$, 
Eqs.~(\ref{eq:fluct}a-b) are written in the compact form
\be
\label{eq:fluct_brief}
\frac{\partial}{\partial t}\R=\mat{M}\R+\vec{\hat{\xi}}\,,
\ee
where $\mat{M}$ is the linear stability matrix of the mean field
solution, and the driving term 
$\vec{\hat{\xi}}=[\hat{\xi},\hat{\xi}^\dagger,\ 0,0]$ includes the
quantum noise of the cavity field. The matrix $\mat{M}$ is non-normal,
therefore it has differrent left and right eigenvectors $\lev{k}$ and
$\rev{k}$, that form a biorthogonal system, {\it i.e.}~their scalar
product $(\lev{k},\rev{l})=\delta_{k,l}$. The quasi-normal modes
defined by $\hat{\rho}_k=(\lev{k},\R)$ are decoupled from each other,
and evolve as
\begin{equation} 
\label{eq:tdfluct}
\hat{\rho}_k(t)=e^{\lambda_k t}\hat{\rho}_k(0)+\int_0^te^{\lambda_k(t-t')}\hat{Q}_k(t') \ud t'\,.
\end{equation}
Generally, the noise enters into all quasi-normal modes via the
projection $\hat{Q}_k = (\lev{k},\vec{\hat{\xi}})$. Since $\R$
contains the fluctuation operators twice (the operators and their
Hermitian adjoint), the operators $\hat{\rho}_k$ also form adjoint
pairs $\rho_+, \rho_+^\dagger$ with eigenvalues $\lambda_{+},
\lambda_{+}^{*}$ and $\rho_-, \rho_-^\dagger$ with eigenvalues
$\lambda_{-}, \lambda_{-}^{*}$. Each pair corresponds to a quasinormal
mode excitation of the system. Fig.~\ref{fig:evalues} shows the
spectrum of the linear stability matrix $\mat{M}$ as a function of
the pumping strength $y$. Solid lines correspond to the imaginary
parts, dashed lines to the real parts of the complex conjugate pairs
of eigenvalues $\lambda_{k}, \lambda_{k}^{*}$.  (The real parts are
the same for each pair.)  The conservative BEC and the lossy cavity
modes form two dissipative quasi-normal excitation modes, both are
subjected to quantum noise and damping. One of them is a photon-like
mode, with eigenvalues $\lambda_{+}$, $\lambda_{+}^{*}$ starting from
$-\kappa \pm{}i\delta_C$ at $y=0$, with a slightly increasing
frequency and decreasing decay rate as $y \rightarrow y_c$. The other
quasi-normal mode dominantly corresponds to the BEC mode. Its
eigenvalues $\lambda_{-}$, $\lambda_{-}^{*}$ are purly imaginary,
$\pm{}i\wR$ at $y=0$, however, with increasing $y$ its decay rate
increases and its frequency decreases down to zero. Interestingly,
there is a finite interval where the imaginary part of $\lambda_{-}$
vanishes.  At the lower and upper limits of this interval, the matrix
$\mat{M}$ becomes defective, i.e. it has only three independent
eigenvectors and three eigenvalues with $\lambda_{-} =
\lambda_{-}^{*}$ becoming a multiple eigenvalue. 
The critical point is reached, where the smallest decay
rate becomes zero. At this point, the quantum noise is not balanced by
damping, therefore the steady-state excitation numbers diverge.

The second order correlations of the
original fluctuation operators can be derived from the correlations
$\left\langle\hat{\rho}_k(t)\hat{\rho}_l(t)\right\rangle$.  In the
regime of cavity cooling, where $\delta_C - u\beta_0^2 < 0$, the real
parts of the eigenvalues $\lambda_k$ are negative, thus the first term
of Eq.~(\ref{eq:tdfluct}) dies out with time. The steady-state
correlations are then obtained from the second term
\be
\label{eq:rhoav}
\left\langle\hat{\rho}_k(t)\hat{\rho}_l(t)\right\rangle \longrightarrow
-\frac{2\kappa}{\lambda_k + \lambda_l}\, {l^{(k)}_1}^*{l^{(l)}_2}^*\,.
\ee
This result is in contrast to the normal mode expectation values of zero temperature systems. For such an equilibrium situation the expectation values are 
simply $\langle\hat{\rho}_k\hat{\rho}_l\rangle\ = 1$, provided that $\hat{\rho}_k$ is the annihilation, and $\hat{\rho}_l$ is the creation operator of the same normal mode, i.e. $0 \leq \;\mathrm{Im}(\lambda_l) = -\;\mathrm{Im}(\lambda_k)$.

The correlations of the original system operators can be calculated using 
their expansion with the quasi normal modes, 
$\R = \sum_k\hat{\rho}_k\rev{k}$, that leads to
\be
\label{eq:sysavr}
\langle{}\hat{R}_i\hat{R}_j\rangle =
\sum_{k,l}\langle\hat{\rho}_k\hat{\rho}_l\rangle\,r^{(k)}_i r^{(l)}_j\,.
\ee
For expample, the condensate depletion is
given by $\delta{}N = \langle\db^\dagger\db\rangle = \langle{}\hat{R}_4\hat{R}_3\rangle$,
while the number of incoherent cavity photons is expressed by 
$\langle\da^\dagger\da\rangle = \langle{}\hat{R}_2\hat{R}_1\rangle$.

\section{Atom-field entanglement}

We quantify the entanglement between the BEC and cavity subsystems by
calculating the logarithmic negativity from the steady-state
correlation matrix. To this end, we introduce the quadrature operators 
$\delta x=(\da+\da^\dagger)/\sqrt{2}$, $\delta{}y=-i(\da -
\da^\dagger)/\sqrt{2}$, $\delta X=(\db+\db^\dagger)/\sqrt{2}$, 
$\delta{}Y=-i(\db - \db^\dagger)/\sqrt{2}$ and group them in the
vector $\vec{u} = (\delta{}x, \delta{}y, \delta{}X,
\delta{}Y)^T$. As these quadratures are hermitian, one can construct a
real correlation matrix
\be
\label{eq:correl_mx}
C_{ij} = \frac12 \langle u_i u_j + u_j u_i \rangle,
\ee 
which has the following block form
\be
\mat{C}=\left[\begin{array}{c c}
\mat{P}&\mat{X}\\
\mat{X}^T&\mat{A}              
\end{array}\right],
\ee where $\mat{P}$ and $\mat{A}$ describes the correlations whithin
the photon and atom fields, while $\mat{X}$ accounts for the cross
correlations between the two. 
The logarithmic negativity can be expressed \cite{Adesso2004Extremal} by the
symplectic invariants ($\det\mat{P}$, $\det\mat{A}$, $\det\mat{X}$) of
the covariance matrix (\ref{eq:correl_mx}) as
\be E_{\cal N} = \max(0, -\log2\tilde{\nu}_{-}), 
\ee 
and
\be \tilde{\nu}_{-} = 2^{-\frac12}\sqrt{\Sigma(\mat{C}) -
  \sqrt{\Sigma(\mat{C})^2 - 4 \det\mat{C}}}, \ee 
where $\Sigma = \det\mat{P} + \det\mat{A} - 2\det\mat{X}$.
The state is separable thus the entanglement is zero if
$\tilde{\nu}_{-}\geq\frac12$. The logarithmic negativity 
quantifies the amount by which this separability criterion is
violated.

\end{document}